\documentclass[twocolumn,showpacs,preprintnumbers,
prb,superscriptaddress]{revtex4}%
\usepackage{amssymb}
\usepackage{amsmath}
\usepackage[dvipdfm]{graphicx} 
\usepackage{dcolumn}
\usepackage{bm}
\usepackage{amsfonts}%
\providecommand{\U}[1]{\protect\rule{.1in}{.1in}}
\ifx\msidvipdf\msiundefined
\else
\fi
\providecommand{\U}[1]{\protect\rule{.1in}{.1in}}

\ifx\pdfoutput\relax\let\pdfoutput=\undefined\fi
\newcount\msipdfoutput
\ifx\pdfoutput\undefined\else
\ifcase\pdfoutput\else
\ifx\paperwidth\undefined\else
\ifdim\paperheight=0pt\relax\else\pdfpageheight\paperheight\fi
\ifdim\paperwidth=0pt\relax\else\pdfpagewidth\paperwidth\fi
\fi\fi\fi
\ifx\msidvipdf\msiundefined\else
\AtBeginDvi{}
\fi
\begin{document}
\title{Dynamical screening in La$_{2}$CuO$_{4}$}
\author{Philipp Werner}
\affiliation{Department of Physics, University of Fribourg, 1700 Fribourg, Switzerland}
\author{Rei Sakuma}
\affiliation{Department of Physics, Division of Mathematical Physics, Lund University,
S\"{o}lvegatan 14A, 223 62 Lund, Sweden}
\author{Fredrik Nilsson}
\affiliation{Department of Physics, Division of Mathematical Physics, Lund University,
S\"{o}lvegatan 14A, 223 62 Lund, Sweden}
\author{Ferdi Aryasetiawan}
\affiliation{Department of Physics, Division of Mathematical Physics, Lund University,
S\"{o}lvegatan 14A, 223 62 Lund, Sweden}
\date{\today }

\begin{abstract}
We show that the dynamical screening of the Coulomb interaction among Cu-$d$
electrons in high-$T_{c}$ cuprates is very strong and that a proper treatment
of this effect is essential for a consistent description of the electronic
structure. In particular, we find that ab-initio calculations for undoped
La$_{2}$CuO$_{4}$ yield an insulator only if the frequency
dependence of the Coulomb interaction is taken into account. We also
identify a collective excitation in the screened interaction at $9$ eV which is
rather localized on the copper site, and which is responsible for a satellite structure
at energy $-13$ eV, located below the $p$ bands.

\end{abstract}

\maketitle


\section{introduction}

The discovery of superconductivity with high transition temperature
$T_{\text{c}}$ in iron pnictide compounds\cite{Kamihara2008} has triggered a
reexamination of the basic theoretical assumptions about the electronic
structure of the copper oxide superconductors,\cite{Sakakibara2010,
Sakakibara2012} based on the similarities and differences between the two
classes of materials. The role and strength of electronic correlations in
high-temperature superconductors is a much debated, but still not completely
settled issue. One of the fundamental problems 
in the theoretical description of a correlated material is the downfolding of
the full many-electron Hamiltonian into a low-energy model with a few orbitals
believed to be most relevant for the origin of superconductivity. In the case
of the cuprates, it is generally agreed that the most relevant orbitals are
those that span the two-dimensional copper oxide layers, namely, Cu
$d_{x^{2}-y^{2}}$ and O $p_{x}$ and $p_{y}$ orbitals, although models that
include the apical O $p_{z}$ as well as Cu $d_{z^{2}}$ orbitals have also been
considered.\cite{Andersen1995} Two prominent low-energy models are the one-band model consisting
of only the strongly hybridized anti-bonding combination of Cu $d_{x^{2}%
-y^{2}}$ and O $p_{x}$ and $p_{y}$ orbitals and a three-band model which also
includes the bonding combination of $d_{x^{2}-y^{2}}$ and $p_{x,y}$ and the
non-bonding $p$ orbital, also known as the Emery model.\cite{Emery1987} Since the stoichiometric compounds are
usually classified as charge-transfer insulators, a proper description of the low-energy properties
should involve the Cu $d_{x^{2}-y^{2}}$ and two oxygen $p_{\sigma}$ orbitals.

A physically well-motivated representation of the underlying one-particle band
structure of these models can be constructed by a tight-binding fit
to the \emph{ab-initio }band structure calculated from the local density
approximation (LDA). Since the Coulomb interaction among the $d$ electrons is
so large that a nonperturbative treatment of the correlation effects is
needed, an interaction term is then added on top of the one-particle
Hamiltonian leading to the Hubbard model with an effective interaction $U$.
The material-specific determination of the Hubbard $U$ is, however, a subtle
and complicated task. It can be shown that a reduction of the Hamiltonian to a
low-energy model necessitates the introduction of a frequency-dependent
$U$ reflecting the retarded electron-electron interaction resulting from the
elimination of the high-energy portion of the original Hamiltonian. In other
words, the frequency-dependent $U$ incorporates the effects of the high-energy
component of the Hamiltonian which has been projected out in the low-energy model.

A large body of 
theoretical studies on the cuprates can
be found in the literature. Several recent works employed a combination of
density functional calculations in the local density approximation and
dynamical mean field theory (LDA+DMFT) in order to investigate one-band and
three-band models of undoped cuprates. The issues discussed in these works are
the importance of antiferromagnetism in opening a gap, or its effect on the
gap size,\cite{Weber2008} the proper choice of the $d$-$p$ level
splitting\cite{Medici2009} and the difference of the electronic structure of 
La$_2$CuO$_4$ in the T and T' crystal structures.\cite{Das2009} It was also shown that the interatomic interaction
between $p$ and $d$ electrons plays an important role in stabilizing the
charge-transfer insulator state, and therefore needs to be considered at least
at the Hartree level.\cite{Hansmann2013} While some of these studies used
realistic bandstructures, the interaction parameters were chosen in an ad-hoc
fashion and as far as we know, all low-energy models for the cuprates
considered so far have neglected the effects of the frequency dependence of
$U$.

The calculation of the Coulomb matrix elements in a Wannier basis
corresponding to the low-energy subspace (here the one-band or  three-band model) is possible
using the constrainded random phase approximation
(cRPA).\cite{Aryasetiawan2004} This formalism yields interaction parameters
which vary from a static value of a few eV (significantly smaller than the
values typically adopted in previous studies) to bare interactions of the
order of 20 eV at high frequency. The importance of properly treating this
frequency dependence has been pointed out in previous
papers,\cite{Werner2010,Werner2012, Huang2012, Casula2013} but not for the cuprates. As
we will show, the screening effect in high-$T_{c}$ cuprates is remarkably 
strong. There are even recent experimental studies which suggest a connection
between screening and $T_{c}$.\cite{Mallett2013}

Four issues will be addressed in the present work: First, what is the role of
the frequency-dependent $U$? Second, is a one-band model sufficient to
describe the low-energy electronic structure of the undoped cuprates? Third,
what is the role of the interaction between the Cu $d$ and O $p$ electrons
that is usually neglected in most studies? 
Fourth, do ab-initio calculations support the conventional classification of
undoped cuprates as charge-transfer insulators?
The third issue has recently been
considered in a model study based on adjustable, static interaction
parameters.\cite{Hansmann2013} 
Here, we focus on the prototypical high-$T_{c}$ material La$_{2}$CuO$_{4}$
which has been thoroughly investigated both experimentally and theoretically.\cite{Damascelli2003}
Our strategy is to perform a \textquotedblleft true" \emph{ab-initio}
simulation of the electronic structure of La$_{2}$CuO$_{4}$, as accurately as
possible with current technology, and to check if it gives a faithful
representation of the low-energy electronic properties. To take into account
electron correlations, we use the DMFT method and solve the impurity problem
with dynamic $U$ using a continuous-time quantum Monte Carlo (CT-QMC)
algorithm. In addition to the \emph{ab-initio} bandstructure we also use the
corresponding \emph{ab-initio }interaction parameters $U$ obtained from the
cRPA method. We find that \emph{ab-initio} calculations which neglect the
frequency dependence of this interaction fail to produce an insulating
solution. On the other hand, if the frequency dependence of the $d$-$d$
interaction is taken into account, a three-band simulation based on \emph{ab-initio}
interaction parameters produces an insulator with a gap size in
good agreement with experiment.

The paper is organized as follows. Section~\ref{method} discusses the methods
used to derive the low-energy models (one-band and three-band) and the LDA+DMFT
approach used to solve these models. Section~\ref{results} shows the spectral
functions obtained for La$_2$CuO$_{4}$ using either the static values of the
estimated Coulomb interactions, or the frequency dependent $d$-$d$
interaction. Section~\ref{summary} is a summary and conclusion.

\section{Model and Method}

\label{method}

\subsection{LDA bandstructure}

\begin{figure}[t]
\begin{center} 
\includegraphics[width=1.0\columnwidth]{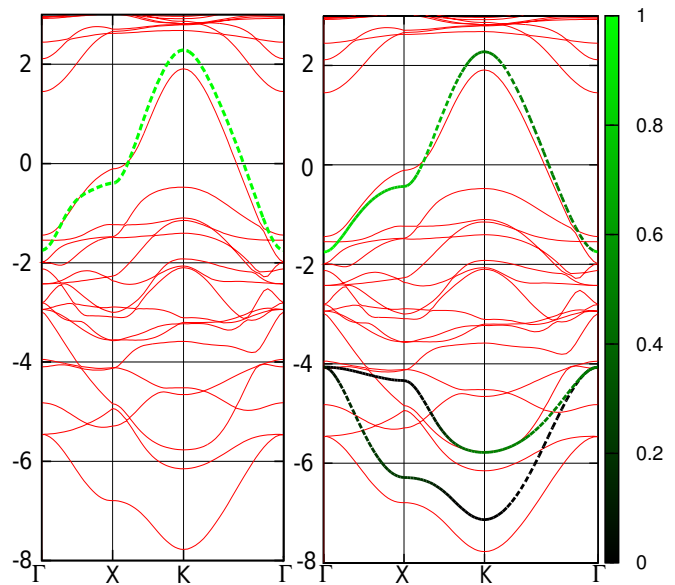}
\end{center}
\caption{LDA bandstructure (solid lines). In addition the left panel shows the
Wannier interpolated band structure for the one-band model and the right panel
the corresponding band structure for the three-band model (thick dashed
lines). The color coding in the right panel indicates the $d$-character of the bands.
The symmetry points are defined as $\Gamma=(0,0)$, $K=(\pi,\pi)$ and $X=(\pi,0)$.
}%
\label{band}%
\end{figure}

Figure \ref{band} shows the LDA bandstructure as well as the bandstructures of
the effective low energy one- and three-band models. The LDA bandstructure was
computed with the full-potential linearized augmented-plane-wave (FLAPW) code
$\mathsf{FLEUR}$ \cite{Fleur} and the model subspaces were defined using
symmetry constrained maximally localized Wannier functions as implemented in
the $\mathsf{WANNIER90}$ library.\cite{Mazari1997,Souza2001,Mostofi2008,Freimuth2008,Sakuma2013a}  
The effective one-band model consists of a single orbital of $d_{x^{2}-y^{2}}$
character at each Cu site. For the three-band model we increase the model
subspace to include also the two in-plane Wannier orbitals of O ${p_{x}/p_{y}%
}$ character. 
It should be noted that, although the conduction bands look very similar in
the two cases the Wannier functions corresponding to the Cu $d_{x^{2}-y^{2}}$
orbitals are very different. In the one-band case the Cu-centred Wannier
function is constructed as a linear combination of only a few bands close to
the Fermi energy. This leads to less variational freedom and hence much more
delocalized Wannier functions than in the three-band case, where more states
are used to construct the Wannier functions. Hence, while in the one-band case 
there is a one to one correspondance between the conduction band and the 
$d_{x^{2}-y^{2}}$-like Wannier function spanning the correlated subspace, 
this is not the case for the three-band model. In the three-band model the 
conduction band can be interpreted as the antibonding combination of the $p$ and $d$ 
states and the two valence bands can be interpreted as the bonding and nonbonding combinations. 
Therefore, although the main $d$-weight is in the conduction band, there is also a 
small $d$-weight in the valence bands as can be seen in the right panel of Fig.~\ref{band}. 

\subsection{cRPA calculation}

We compute the frequency-dependent interaction parameters for the one-band and
three-band model using cRPA.\cite{Aryasetiawan2004} In this scheme, the polarization function
$P(\omega)$ is calculated in the random phase approximation, i.e. by
considering only the bubble diagrams with LDA propagators. This polarization
function relates the bare interaction $V$ and the fully screened Coulomb
interaction $W(\omega)$ via%

\[
W(\omega)=V+VP(\omega)W(\omega).
\]
One then defines a polarization $P_{l}(\omega)$ associated with transitions
between states defined in the low-energy bands. Since these transitions will
be treated explicitly in the DMFT calculation, we remove these screening
processes in the calculation of the effective interaction. For this, one
computes $P_{r}(\omega)=P(\omega)-P_{l}(\omega)$ and the frequency-dependent
$U(\omega)$ which satisfies%

\[
W(\omega)=U(\omega)+U(\omega)P_{l}(\omega)W(\omega).
\]
Apparently, the screening of $U(\omega)$ by $P_{l}(\omega)$ gives the fully screened
interaction $W(\omega)$. One thus interprets $U(\omega)$ as an effective
frequency-dependent interaction among electrons residing in the low energy
subspace defining the Hilbert space of the low-energy model. This so-called
Hubbard $U$ can also be obtained by solving the following equation:%
\begin{align}
&  U(\mathbf{r,r}^{\prime};\omega)=V(\mathbf{r,r}^{\prime})\nonumber\\
&  \hspace{5mm}+\int d^{3}r_{1}d^{3}r_{2}V(\mathbf{r,r}_{1})P_{r}%
(\mathbf{r}_{1},\mathbf{r}_{2};\omega)U(\mathbf{r}_{2},\mathbf{r}^{\prime
};\omega), \label{U}%
\end{align}
or schematically $U(\omega)=[1-VP_{r}(\omega)]^{-1}V$. The frequency-dependent
interaction parameters of the model are then given by the matrix elements of
$U(\omega)$ in the Wannier basis $\left\{  \varphi_{m}\right\}  $ constructed
using the procedure of Marzari and Vanderbilt:\cite{Mazari1997}

\begin{align}
&  \left\langle \varphi_{m_{1}}\varphi_{m_{2}}|U(\omega)|\varphi_{m_{3}%
}\varphi_{m_{4}}\right\rangle =\int d^{3}rd^{3}r^{\prime}\varphi_{m_{1}}%
^{\ast}(\mathbf{r)}\varphi_{m2}(\mathbf{r)}\nonumber\\
&  \hspace{30mm}\times U(\mathbf{r,r}^{\prime};\omega)\varphi_{m_{3}%
}(\mathbf{r}^{\prime}\mathbf{)}\varphi_{m_{4}}^{\ast}(\mathbf{r}^{\prime
}\mathbf{).} \label{Umat}%
\end{align}

While the application of this procedure to the one-band model is unambiguous,
the three-band case is more subtle. Here, the subset of screening
processes which should be excluded depends on how the three-band model is
solved. If we were to solve the full three-band model, we would simply remove
all screening processes within the model and no ambiguity would arise. It
would, however, lead to a multi-site impurity problem involving not only the
copper site but also the oxygen sites and orbital-dependent $U(\omega)$. At
present it is not possible to perform DMFT calculations for such a complex
problem. In this work, we will treat the $d$-$d$ interactions within DMFT, and
the $p$-$p$ and $p$-$d$ interactions at the Hartree level (similar to
Ref.~\onlinecite{Hansmann2013}). In this case, only the $d$-$d$ screening
needs to be removed in the calculation of $U$ since we do not include $p$-$d$
screening processes in the model. 
According to the discussion in the previous section,
the main $d$-weight is in the conduction band. We therefore remove only the screening
within the conduction band also for the three-band model.
The effective interaction $U(\mathbf{r,r}%
^{\prime};\omega)$ as defined in Eq.~(\ref{U}) is then the same as in the
one-band model but the matrix elements of $U$ as defined in Eq.~(\ref{Umat})
representing the interaction between $d$-electrons will nevertheless be
different from the one-band case, because the Wannier orbitals of the
three-band model are significantly more localized.

\subsection{DMFT calculation}

The LDA calculation and cRPA downfolding lead to a low energy effective model
with one or three bands and dynamically screened (retarded) intra- and
inter-orbital interactions. To solve this model, we use the DMFT
method.\cite{Georges1996} This approximation maps the lattice problem onto a
single-orbital Anderson impurity model with a dynamical interaction
$U_{dd}(\omega)$, i.e. an electron-boson problem with a Holstein-like coupling
to a continuum of bosonic modes.\cite{Werner2010} Using the
hybridization-expansion Monte Carlo method,\cite{Werner2006, Werner2007} this
impurity problem can be solved efficiently and without approximations on the
imaginary axis, yielding the impurity Green's function $G_{\text{imp}}%
(i\omega_{n})$ and the impurity self-energy $\Sigma_{\text{imp}}(i\omega_{n})$.

In the one-band model, we approximate the lattice self-energy $\Sigma
(k,i\omega_{n})$ by $\Sigma_{\text{imp}}(i\omega_{n})$ and compute the local
lattice Green's function as%

\[
G_{\text{loc}}(i\omega_{n})=\int(dk)[i\omega_{n}+\mu-\varepsilon_{k}%
-\Sigma_{\text{imp}}(i\omega_{n})]^{-1}.
\]
Here, the $k$-integral is normalized over the Brillouin zone, and
$\epsilon_{k}$ is the conduction band dispersion. The chemical potential $\mu$ is
adjusted to ensure one $d$-electron per unit cell, so we do not need a
\textquotedblleft double counting term" to remove the Hartree-type
self-energy contribution which is already included at the LDA level.

The three-band case needs some justification. Let us start with the
Hamiltonian with a static $U=U(\omega=0)$ given by%

\begin{align}
H=H_{0}  &  +U_{dd}\sum_{i}n_{id\uparrow}n_{id\downarrow}+U_{pp}\sum
_{j}n_{jp\uparrow}n_{jp\downarrow}\nonumber\\
&  +U_{pd}\sum_{\left\langle ij\right\rangle }n_{id}n_{jp},
\end{align}
where $n=n_{\uparrow}+n_{\downarrow}$, $H_{0}$ is the tight-binding
Hamiltonian for the three-band model and $i$ and $j$ label the copper and
oxygen sites, respectively. Since the $p$ bands are filled, correlation
effects among $p$ electrons are expected to be small and the LDA bands should
be quite reliable. The impurity problem is therefore solved only for the
copper site and since we do not consider $p$ to $d$ screening channels in the
model, the effective interaction $U_{dd}$ must include these $p$-$d$ screening
processes and therefore corresponds to the one-band model, albeit evaluated with the more localized Wannier
functions of the three-band model, as discussed earlier. We now
take into account the frequency dependence of $U_{dd}$ and solve the impurity
problem with a dynamic $U_{dd}$ using the CT-QMC method within the action
formalism. In the three-band case, we consider, in addition to the local
self-energy $\Sigma_{dd}(i\omega_{n})=\Sigma_{\text{imp}}(i\omega_{n})$ the
$p$-$p$ and $p$-$d$ interactions at the Hartree level. We thus have to add
double counting terms $\Sigma_{DC}$, which as in
Ref.~\onlinecite{Hansmann2013} we evaluate with the LDA densities for the
$U_{pp}$ and $U_{pd}$ contributions. 
This amounts to adjusting the Hartree self-energies (which are included in the
LDA) to the self-consistently computed densities. For $\Sigma_{dd}$, we use a
standard double-counting term\cite{Anisimov1991} evaluated with the correlated
density $n_{d}$.\cite{DC} Specifically, the diagonal matrix elements of
$\tilde\Sigma=\Sigma-\Sigma_{DC}$ are
\begin{align}
\tilde\Sigma_{dd}(i\omega_{n})  &  =\Sigma_{\text{imp}}(i\omega_{n}%
)-U_{dd}(0)(n_{d}-\tfrac{1}{2})\nonumber\\
&  \hspace{20mm}+4U_{pd}(0)(n_{p}-n_{p}^{LDA}),\label{hartree1}\\
\tilde\Sigma_{pp}(i\omega_{n})  &  =U_{pp}(0)(n_{p}-n_{p}^{LDA})+2U_{pd}%
(0)(n_{d}-n_{d}^{LDA}),\label{hartree2}
\end{align}
and the off-diagonal elements are set to zero. The factor of four in the last
term of $\tilde\Sigma_{dd}$ is due to the presence of four nearest oxygen
atoms around a copper atom and the factor of two in the last term of
$\tilde\Sigma_{pp}$ is due to the presence of two nearest copper atoms around
an oxygen atom.\ Note that in the Hartree-like terms, we use the screened
interactions. While this can be justified in the case of the $d$-$d$
interaction,\cite{Werner2012} it is an approximation for the $U_{pp}$ and $U_{pd}$ terms which
should be considered as a lower bound estimate. At present, it is unclear how
the frequency-dependence should be incorporated into a static description if
the screening modes for different interation terms are different.

With this approximate self-energy, we then compute the local lattice Green's
function as%

\[
G_{\text{loc}}(i\omega_{n})=\int(dk)[(i\omega_{n}+\mu)\mathcal{I}-H_{k}%
-\tilde{\Sigma}(i\omega_{n})]^{-1}%
\]
which is a $3\times3$ matrix, and then extract the $d$-component in order to
define a new hybridization function for the impurity model. In the
self-consistent iteration, the chemical potential is adjusted such that the
total number of $p$- and $d$-electrons is $\sum_{\alpha=1}^{3}G_{\alpha\alpha
}(\tau=0_{-})=5$.

\subsection{Analytical continuation}
\label{maxent}

In order to compute spectral functions for models with frequency dependent interactions, one can 
use the strategy proposed in Ref.~\onlinecite{Casula2012}. We define the
bosonic function $\exp[-K(\tau)]$, with%

\[
K(\tau)=\frac{1}{\pi}\int_{0}^{\infty}d\omega^{\prime}\frac{\text{Im}%
U(\omega^{\prime})}{\omega^{\prime2}}[b(\omega^{\prime},\tau)-b(\omega
^{\prime},0)]
\]
and 
$b(\omega^{\prime},\tau)=\cosh[(\tau-\beta/2)\omega^{\prime}]/\sinh[\beta
\omega^{\prime}/2]$, 
and compute the auxiliary Green's function $G_{\text{aux}}(\tau)=G_{dd}%
(\tau)/\exp[-K(\tau)]$. The spectral function corresponding to $G_{\text{aux}%
}(\tau)$ is expected to have no high-frequency components and can be obtained
using the maximum entropy analytical continuation procedure.\cite{Jarrell1996}
Finally, the spectral function $A$ for $G$ is obtained from a convolution of
the auxiliary spectral function $A_{\text{aux}}$ and the exactly computable
spectral function of the bosonic factor $\exp[-K(\tau)]$.\cite{Casula2012,
Werner2012} In this convolution, the low-energy structures of the spectral
function are replicated at energies which are directly related to the dominant
screening modes.

We can employ the same strategy to analytically continue the self-energy. For
this, we first compute a Green's function $\tilde G(i\omega_{n})=1/(i\omega
_{n}+\tilde\mu-\Sigma(i\omega_{n}))$ with a suitably chosen $\tilde\mu$ and
apply the above procedure to obtain the corresponding spectral function
$\tilde A(\omega)$ and (using the Kramers-Kronig transformation) the Green's
function $\tilde G(\omega)$. The real-frequency self-energy, including
high-energy features, is then given by $\Sigma(\omega)=\omega+\tilde
\mu-1/G(\omega)$.

\begin{figure}[t]
\begin{center}
\includegraphics[angle=-90, width=\columnwidth]{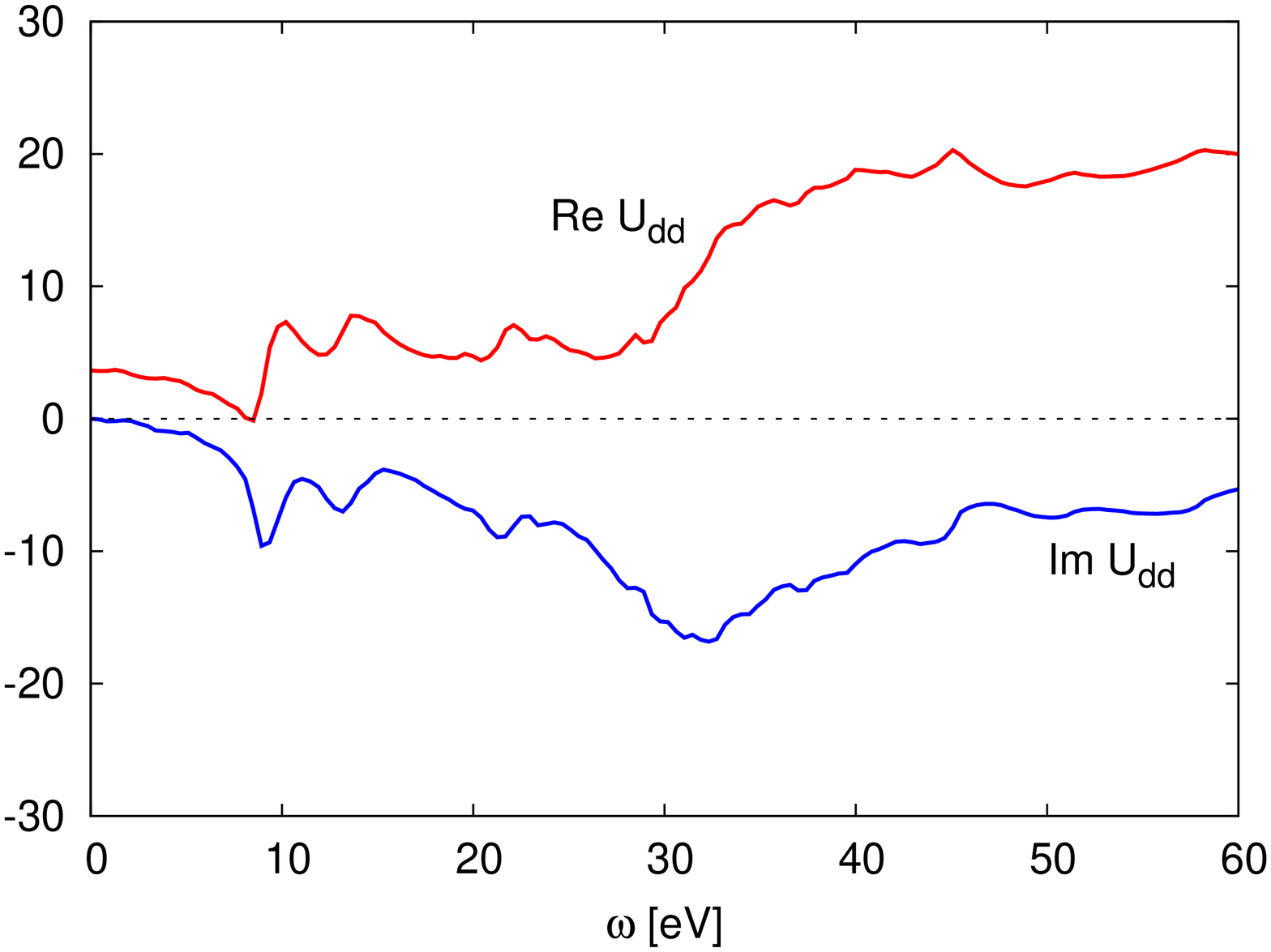}\newline%
\includegraphics[angle=-90, width=\columnwidth]{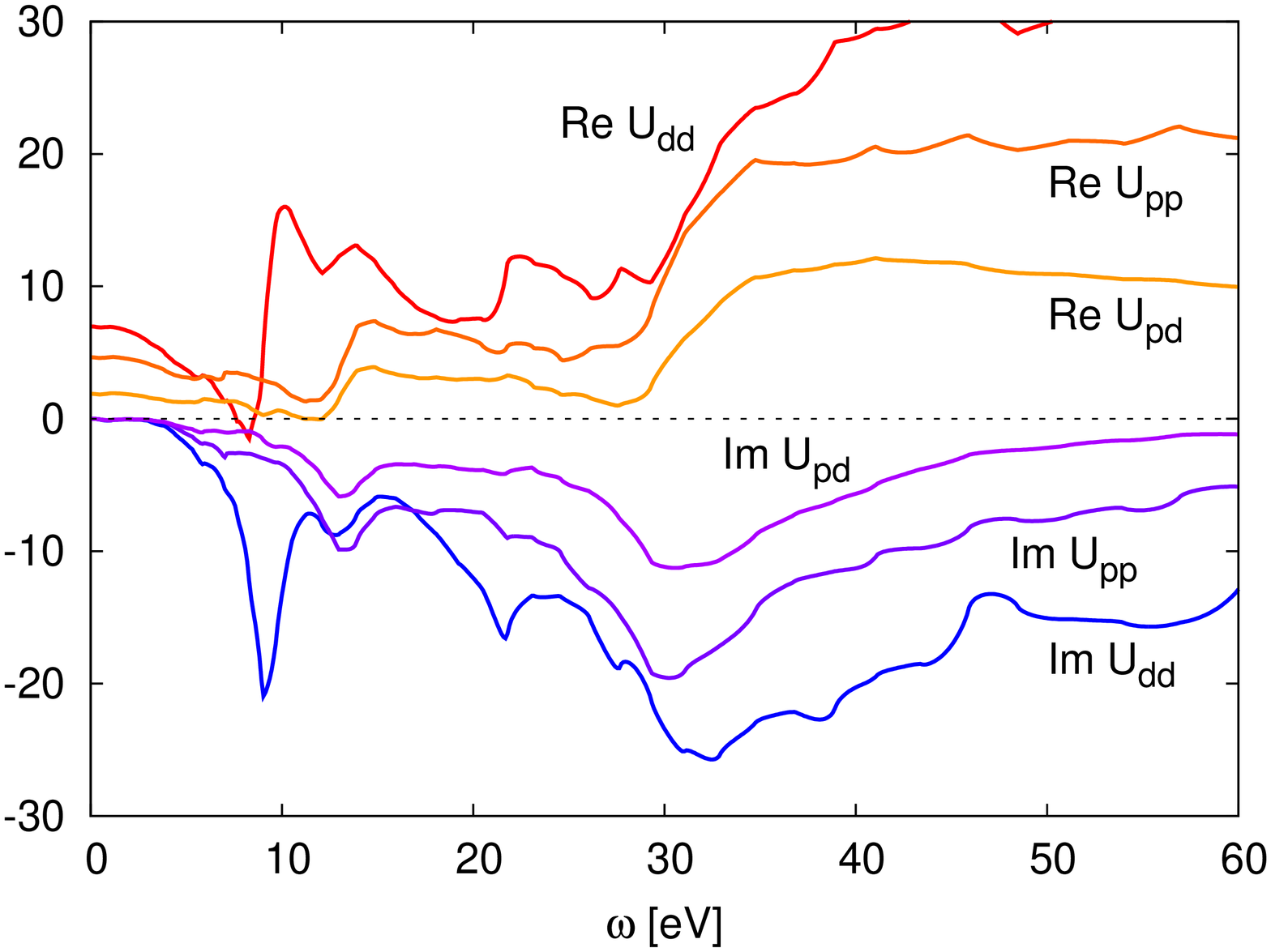}
\end{center}
\caption{Dynamically screened interactions in the one-band model (top panel) and
in the three-band model (bottom panel). Because only the $d$-orbital is considered
in the impurity calculation, we only remove the $d$-$d$ screening in the
three-orbital case.}%
\label{interactions}%
\end{figure}

\section{Results}

\label{results}

\subsection{Frequency-dependent $U$}

We plot the cRPA results for La$_{2}$CuO$_{4}$ in Fig.~\ref{interactions}. The
top panel shows $U_{dd}(\omega)$ for the one-band model and the bottom panel
shows $U_{dd}(\omega)$, $U_{pp}(\omega)$ and $U_{pd}(\omega)$ for the
three-band model. In the one-band case, the static (screened) interaction is
$U_{dd}(\omega=0)=3.65$ eV. The imaginary part of $U$, which describes the
excitation spectrum of the system excluding contributions from the model, is
characterized by several collective excitations: a broad peak centered at
$\omega=30$ eV and a sharp peak at $\omega=9$ eV, as well as smaller peaks
around $\omega=13$ and $21$ eV. The broad peak corresponds to a collective
plasmon excitation that is coupled to single-particle excitations providing
decaying channels responsible for the broad feature. The pole-like structure
around $\omega=9$ eV may be interpreted as a collective subplasmon excitation
arising from single-particle transitions from the occupied oxygen $p$ bands to
the unoccupied part of the anti-bonding $d_{x^{2}-y^{2}}$ band. At very
high-energy above the plasmon frequency, screening becomes ineffective and the
interaction approaches the bare Coulomb interaction value of $U_{dd}\approx20$ eV.

In the three-band case, the structures of the frequency-dependent $U_{dd}$
interaction look similar to the one-band case but the static value is
$U_{dd}(\omega=0)=7.00$ eV, while the high frequency limit is about 30 eV.
These higher values result from the more localized Wannier orbitals because,
as explained previously, $U_{dd}$ is calculated as a matrix element of the
$U(\mathbf{r,r}^{\prime};\omega)$ of the one-band model. The static values of
the $p$-$p$ and $p$-$d$ interactions are $U_{pp}(\omega=0)=4.64$ eV and
$U_{pd}(\omega=0)=1.88$. For these interactions, the dominant low-frequency
pole is near $13$ eV (the peak at $9$ eV is missing). 
Since $U_{dd}\,$, $\ U_{pp}$, and
$U_{pd}$ are calculated as matrix elements of the \emph{same} $U(\mathbf{r,r}%
^{\prime};\omega)$, the presence of a strong peak in $\operatorname{Im}U_{dd}$
but not in $\operatorname{Im}U_{pp}$ and $\operatorname{Im}U_{pd}$ 
implies that the collective excitation corresponding to the $9$ eV peak is not
extended, as in usual plasmon-like excitations, but rather localized on the
copper site. This suggests that the screening mechanism of a hole or a test
charge created at the copper site will be rather different from the screening mechanism at the
oxygen sites. An additional screening charge fluctuation associated with the
$9$ eV peak is present in response to a hole created at the copper site but
not at the oxygen sites. For HgBa$_2$CuO$_4$, another high-$T_c$ cuprate compound, one can identify the same low
frequency features around $9$ eV in $U_{dd}$, while the corresponding feature is absent in both $U_{pd}$ and $U_{dd}$. 
This indicates that the localized $p$-$d$ excitation at $9$ eV might 
 be a universal feature of the cuprate compounds.

A useful way to quantify the screening effect is to compute the
\textquotedblleft renormalization factor"\cite{Casula2013} $Z_{B}=\exp
[\frac{1}{\pi}\int_{0}^{\infty}\text{Im}U(\omega)/\omega]$. In a one-band
model, the low-energy properties of the solution for a frequency-dependent
interaction $U(\omega)$ can be reproduced by a calculation involving the
static interaction $U(\omega=0)$ and hopping parameters renormalized by
$Z_{B}$. Hence, this factor essentially tells us by how much the static limit
underestimates the interaction strength. In the case of La$_{2}$CuO$_{4}$,
$Z_{B}$ is remarkably low. For the one-band model, we find $Z_{B}=0.58$ and
for the three-band model $Z_{B}^{dd}=0.52$. These low values are primarily due
to the strong pole near $9$ eV. Indeed, for $U_{pp}$ and $U_{pd}$ the
renormalization factor is higher: $Z_{B}^{pp}=0.68$ and $Z_{B}^{pd}=0.80$.
$Z_{B}$ is low for La$_2$CuO$_4$ even in comparison with other cuprate compounds. 
For HgBa$_2$CuO$_4$ for example, where the $9$ eV pole is less pronounced, the renormalization factor is $Z_{B}=0.66$ in the one-band model.

\begin{figure}[t]
\begin{center}
\includegraphics[angle=-90, width=\columnwidth]{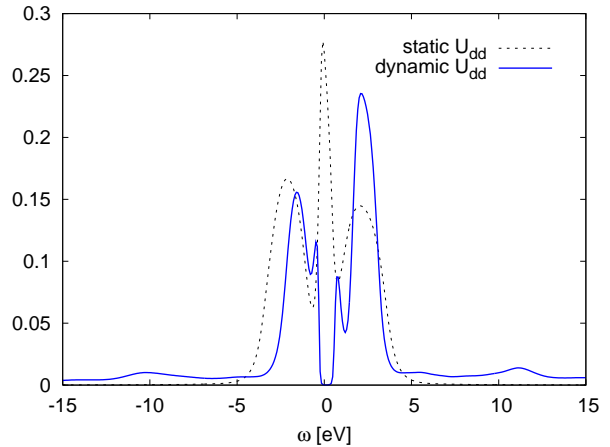}
\end{center}
\caption{Local spectral functions for the one-band model with static and dynamic $U$ (inverse
temperature $\beta=10$). Results obtained via analytical continuation of the
self-energy.}%
\label{dos_oneband}%
\end{figure}

\begin{figure}[t]
\begin{center} 
\includegraphics[angle=-90, width=\columnwidth]{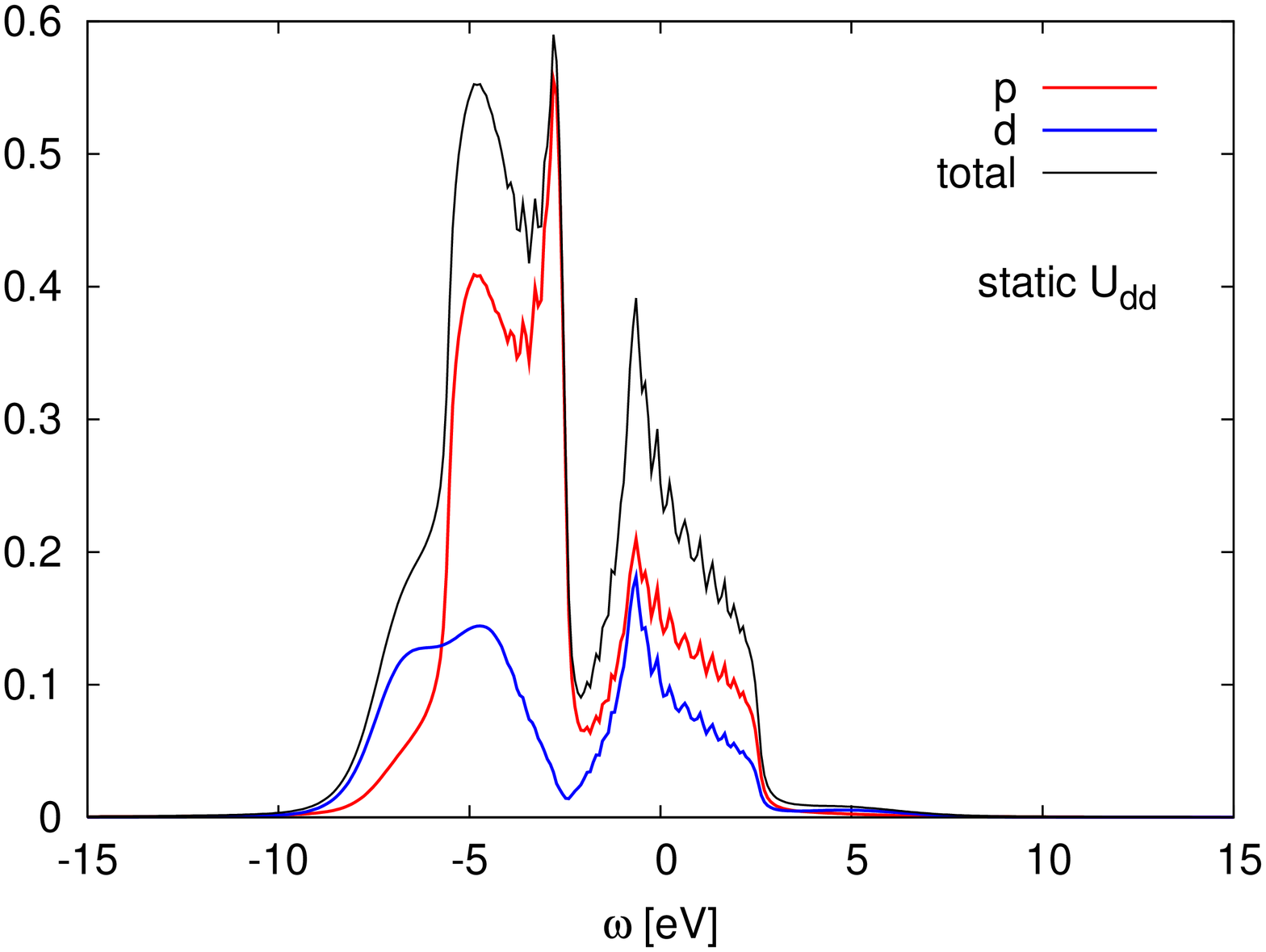}
\includegraphics[angle=-90, width=\columnwidth]{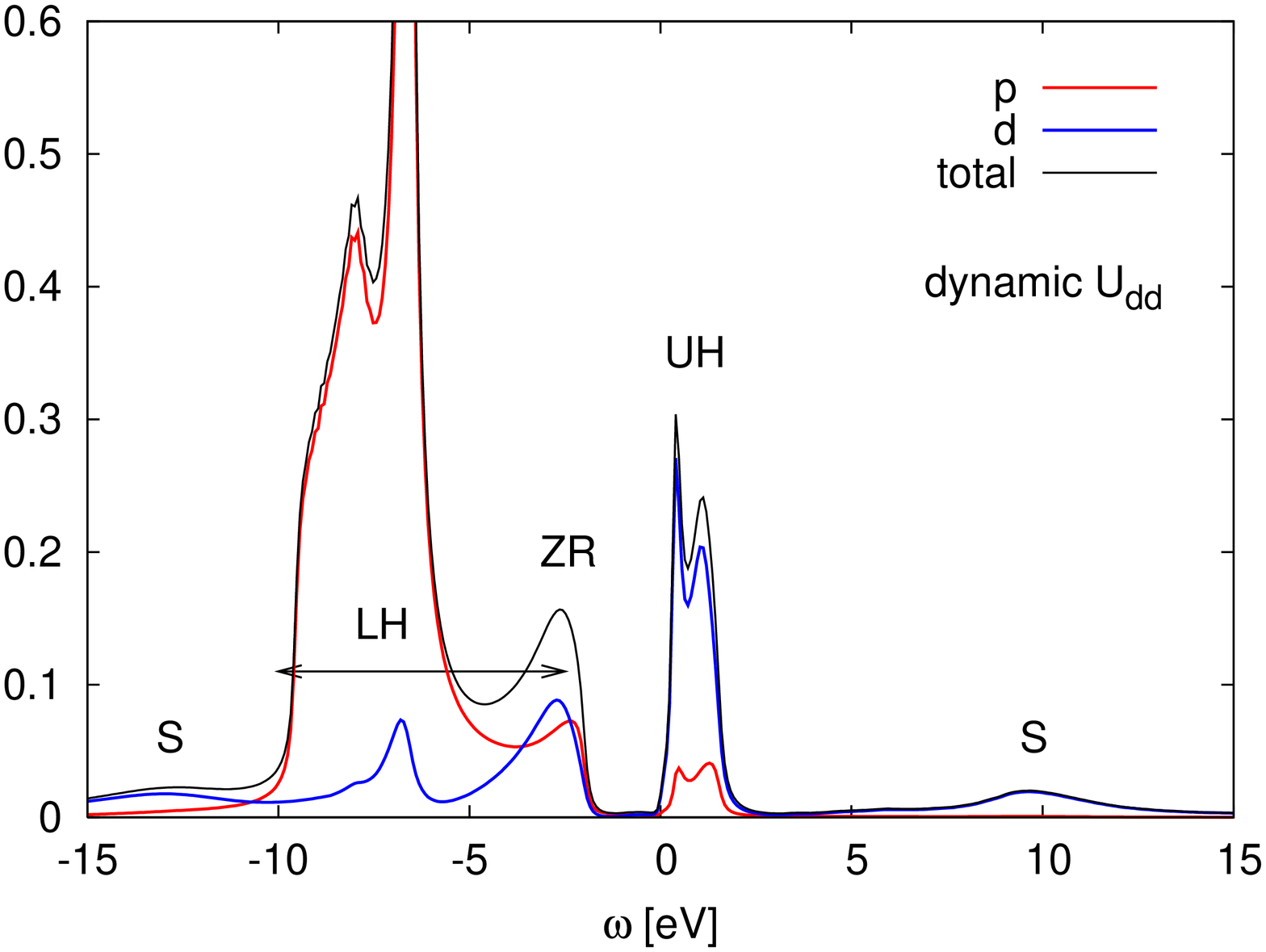}
\end{center}
\caption{Local spectral functions for the three-band model with static $U$ (top panel) 
and with dynamic $U$ (bottom panel) at inverse temperature $\beta=10$. 
We have identified the following features in the $d$ spectral function: upper Hubbard band (UH), lower Hubbard band (LH), Zhang-Rice singlet band (ZR), and satellites (S). 
Results obtained via analytical continuation of the
self-energy. }%
\label{dos}%
\end{figure}

\subsection{One-band model}

\label{sec_oneband}

We first discuss the results obtained for the one-band model. 
Figure~\ref{dos_oneband} shows the local $d$-electron spectral function obtained with
the frequency-dependent interaction (blue line) and with the static
interaction $U_{dd}(0)$ (dashed black curve). The calculations have been performed
at temperature $T=0.1$ in the paramagnetic phase, and we use the analytical continuation 
procedure described in Sec.~\ref{maxent}. We see that the static
interaction is not enough to open a Mott gap in the spectral function, whereas
the calculation with the full $U_{dd}(\omega)$ yields a gap. However, the
gapsize of $\lesssim1$ eV is too small compared to the experimentally measured
optical gap of 2 eV.\cite{Ginder1988} Apart from this low-energy region, the
spectra differ mainly at high energies. Here, the dynamic-$U$ spectrum
features satellites at energies of approximately $\pm9$-$13$ eV and a broad
plasmon peak centered around $30$ eV. They correspond to collective exciations
with simultaneous emission or absorption of quantized density fluctuations
with a frequency given by the dominant modes visible in Fig.~\ref{interactions}. 
Obviously, this physics is missing in a static-$U$ description.

We have also performed a calculation with a static $U$ but with the one-particle
band renormalized by the Bose factor
$Z_{B}$ as proposed in Ref.~\onlinecite{Casula2013}. This calculation also produces a gap,
confirming the importance of the frequency dependence of $U$ in renormalizing the band width.

\subsection{Three-band model}

In the three-band calculations, we also find that static interactions equal to
the static limit of the {\it ab-initio} estimated interaction parameters are not
enough to open a gap in the spectral function (upper panel of Fig.~\ref{dos}). However, as shown in the bottom
panel of Fig.~\ref{dos}, if the frequency-dependence of $U_{dd}$ is
considered, an insulating solution is found, with a gap of $1.9$ eV. This is in
rather good agreement with the experimentally measured gap. (Our use of 
the static screened interactions in the Hartree terms
implies that this calculation yields a lower bound for the gap size.)
Furthermore, since the calculations have been performed in the
paramagnetic phase, the gap opening confirms that the insulating nature of
La$_{2}$CuO$_{4}$ is of Mott-Hubbard rather than Slater type.
We also find, in agreement with Ref.~\onlinecite{Hansmann2013}, that the interatomic Hartree potential is
essential: Without the corresponding shift in the relative $p$-$d$ level splitting, the frequency-dependent $U_{dd}$
would not be enough to open a Mott gap.

In contrast to the one-band result, the $d$-spectral function from the
three-band calculation is strongly asymmetric, due to the hybridization with
the $p$-states which lie below the Fermi energy. The states near the lower gap
edge 
have a mixed $p$-$d$ character
($d^{8}$ ligand hole) and correspond to the \textquotedblleft Zhang-Rice" 
singlet band.\cite{Zhang1988, Medici2009} On the unoccupied side, the $d$ density of
states is peaked near the band edge and extends over an energy range of about
2 eV. This feature may be interpreted as the upper Hubbard band corresponding to the
$d^{10}$ configuration.

While the upper Hubbard band is rather well defined, there have been
conflicting results concerning the lower Hubbard band. In view of the
discussion in the previous literature about the correct position of the lower
Hubbard band\cite{Medici2009, Weber2008} we have to caution that this feature
is difficult to identify due to the dynamical nature of the Coulomb
interaction. Especially in La$_{2}$CuO$_{4}$, which has a prominent screening
mode at $\omega\approx9$ eV (similar to the screened interaction of
$U_{dd}(\omega=0)=7$ eV) structures that may be identified with
the lower Hubbard band 
can be expected to overlap with satellite features.
Furthermore, due to the self-consistent adjustment of the $p$-$d$ level splitting via the Hartree contribution in 
Eqs.~(\ref{hartree1}) and (\ref{hartree2}), 
which is affected by the smaller $n_d$ in the dynamic-$U$ calculation, the $p$-states are pushed down in energy, 
so that there is a strong $p$-$d$ hybridization in the energy range
where we expect the lower Hubbard band. 

\begin{figure}[t]
\begin{center}
\includegraphics[angle=-90, width=\columnwidth]{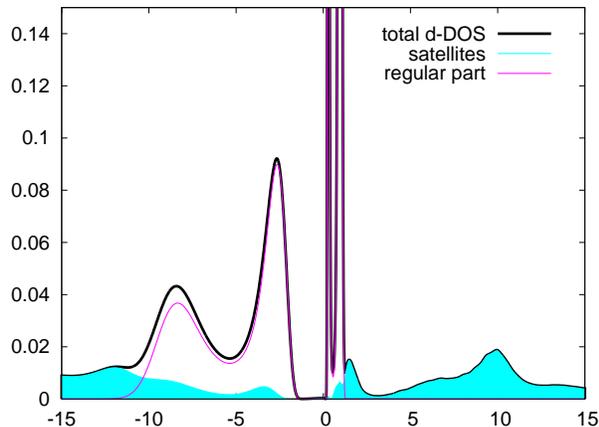}
\end{center}
\caption{Local $d$-electron spectral function for the 3-band model (inverse
temperature $\beta=10$).
Results obtained via analytical continuation of the local Green's function. The black line plots the total spectral function,
while the pink line shows $A_\text{reg}(\omega)$ and the blue shaded area the satellite contributions 
generated by the convolution of $A_\text{reg}(\omega)$ with the bosonic spectral function.
}%
\label{dos_fromG}%
\end{figure}

\begin{figure}[t]
\begin{center} 
\includegraphics[angle=-90, width=\columnwidth]{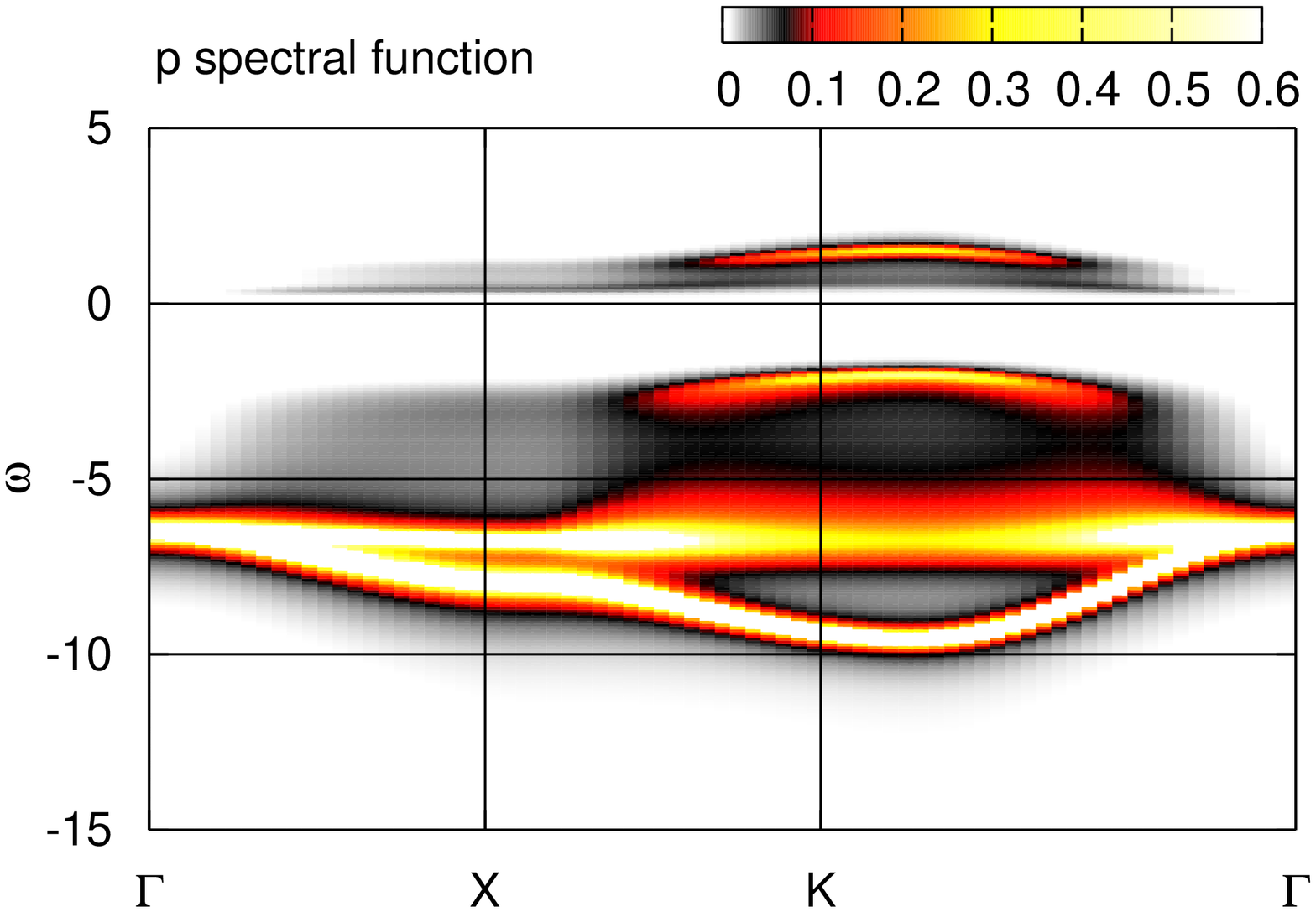}
\includegraphics[angle=-90, width=\columnwidth]{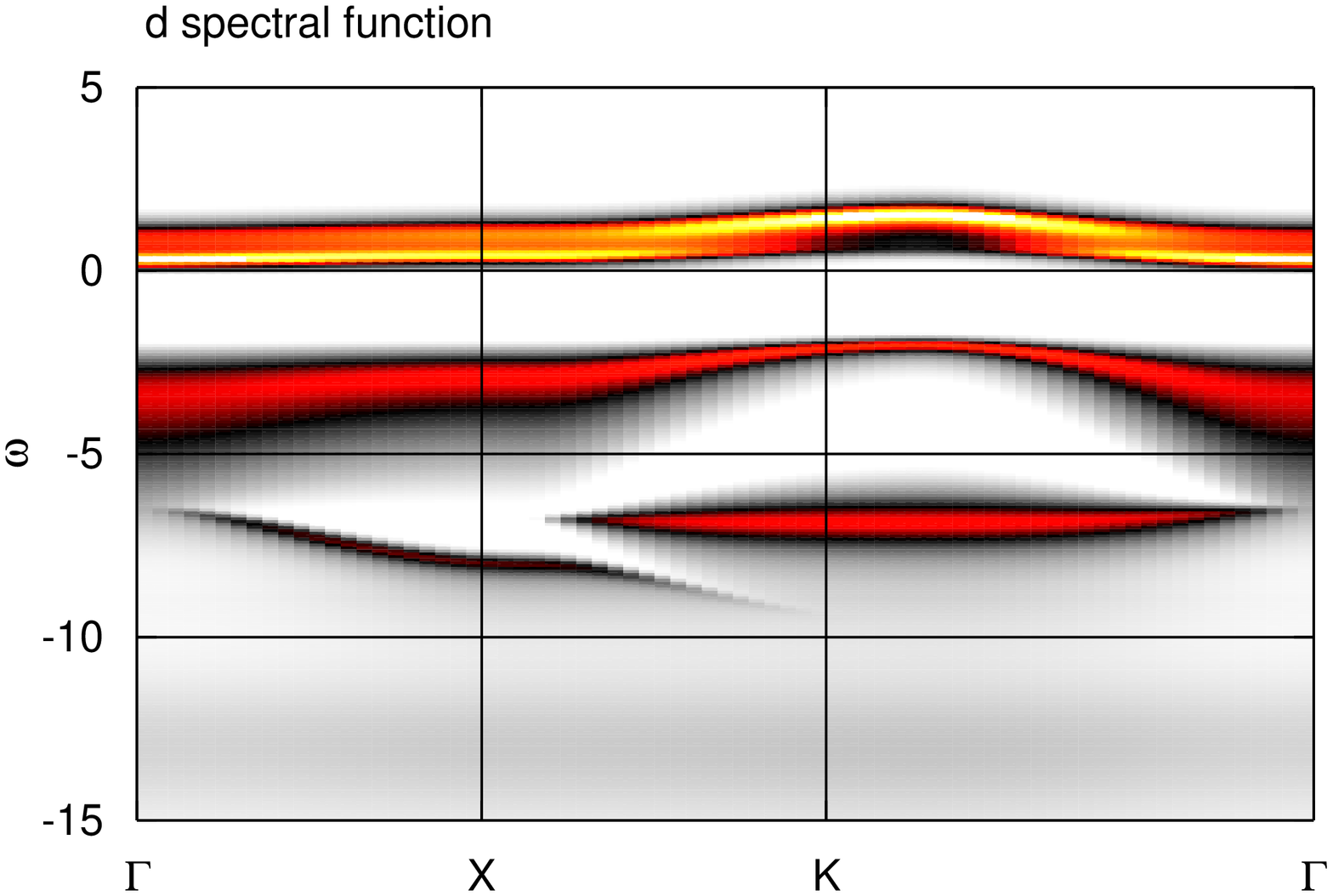}
\end{center}
\caption{$k$-resolved spectral functions for the three band model at $\beta=10$ along the same path as in Fig. 1. Top panel: $p$-electron spectral function. Bottom panel: $d$-electron spectral function.  
}%
\label{dos_band}%
\end{figure}

To shed some light on the satellite issue, we plot in Fig.~\ref{dos_fromG} the
$d$-electron spectral function for the three-band model. In contrast to
Fig.~\ref{dos}, where the density of states has been obtained via the
analytical continuation of the self-energy, we computed the spectral function
shown in Fig.~\ref{dos_fromG} directly from the local Green's function, by the
procedure explained in Sec.~\ref{maxent}. While the
direct continuation of the Green's function yields a somewhat poorer
resolution of the features in the energy region dominated by the $p$-states,
the agreement between the two spectral functions is rather good.

The analytical continuation of $G$ by the method of Casula {\it et
al.}\cite{Casula2012} allows us to identify a ``regular" contribution to the
density of states, and a ``satellite" contribution, corresponding to states
which can be accessed via the emission or absorption of bosons. In the regular
part, we can identify the upper Hubbard band in the energy region from 0 to 2
eV, the Zhang-Rice singlet band responsible for the peak near the lower gap edge,
and a broad feature in the energy range from -3 to -10 eV. It is this latter
feature which should be associated with the lower Hubbard band. The comparison
with Fig.~\ref{dos} shows that this Hubbard band (which is somewhat more
asymmetric in the spectrum based on the analytical continuation of $\Sigma$)
overlaps with the $p$ states, so that the lower Hubbard band is partially 
masked by $d$ spectral weight originating from $p$-$d$ hybridization. 
On the other hand, the hump seen in the energy range from -10 to
-15 eV, as well as the peak centered around +10 eV, should be considered
satellite features which result from the frequency dependence of the
Hubbard-$U$ in the effective low-energy model. The position of the 
satallite feature around -13 eV is in good agreement with the experimental photoemission
spectra in Ref.~\onlinecite{Shen1987}.

\begin{figure}[t]
\begin{center}
\includegraphics[angle=-90, width=\columnwidth]{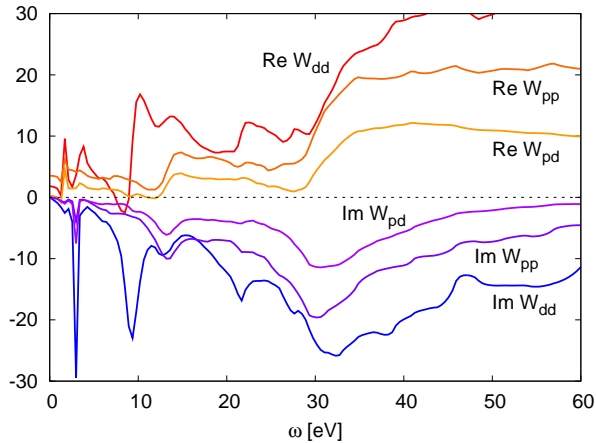}
\end{center}
\caption{Fully screened interaction $W(\omega)$ for the three-band model. 
}%
\label{W}%
\end{figure}

To reveal the lower Hubbard band, it is instructive to look at the momentum resolved spectral function. In Fig.~\ref{dos_band}, 
we plot the $p$ and $d$ spectral functions along the same path as in Fig.~\ref{band}. Besides the weakly dispersing upper Hubbard 
band we find a similarly dispersing band in the energy range from $-2$ to $-6$ eV. The states near the band edge, which have 
a strong overlap with $p$ states, may be identified with Zhang-Rice singlets. 
In the same region of momentum space, one finds an almost dispersionless band at $-7$ eV, 
which also exhibits a strong overlap with $p$ states. This energy is suggestive because it corresponds to the screened $U_{dd}$, 
and the chemical potential is at the upper gap edge. However, a comparison with the LDA bandstructure in Fig.~\ref{band} 
and the $p$ spectral function shows that this feature in the $d$ spectral function can be naturally interpreted as originating from 
the hybridization with a renormalized $p$ band. Hence, it appears that the $d$ states which may be associated with 
the lower Hubbard band cover a broad energy range up to the gap edge, and that the Zhang-Rice band should be considered 
a substructure of the lower Hubbard band.   
Since the lower Hubbard band covers the same energy range as the $p$ states, and the structure near $-13$ eV 
(visible as a grey band in the bottom panel of Fig.~\ref{dos_band}) is a satellite, our calculation is not consistent with a simple 
charge-transfer insulator picture, in which the Hubbard band lies below the $p$-states.

It is interesting to note that the fully screened interaction $\operatorname{Im}W_{dd}$ is dominated by two strong peaks at
energies $3$ and $9$ eV which signal the formation of many-body or collective
states with those binding energies (Fig.~\ref{W}). Comparison with $\operatorname{Im}U_{dd}$
for the one-band or three-band model allows us to conclude that the peak at
$3$ eV in $\operatorname{Im}W_{dd}$ originates from collective excitations
within the $d_{x^{2}-y^{2}}$ band since the peak is missing in
$\operatorname{Im}U_{dd}$. This energy happens to be close to the size of the gap. 
In a weakly correlated system structures in $\operatorname{Im}W$ must necessarily be carried over to $\operatorname{Im}\Sigma$ 
and in turn
inherited by the spectral function. Structures in the spectral function must
therefore reflect structures in $\operatorname{Im}W$. One interesting but yet
unresolved issue is the relation between these peaks in $\operatorname{Im}W$ and the Hubbard bands.

If the screened interaction $W$ were computed fully self-consistently, and not by cRPA, transitions across the gap 
would contribute to the low-energy screening, 
so that we can expect a feature in $\text{Im}W_{dd}$ at an energy corresponding to this gap. 
There is however a priori no reason why the cRPA $W$, which is derived from the 
LDA bandstructure, should exhibit these structures. 
Whether the agreement between the gap size in Fig.~\ref{dos_fromG} and the sharp peak in $\text{Im}W$ in Fig.~\ref{W}
is a mere conincidence, or if the corresponding properties of the bandstructure (used in the DMFT calculation) 
play a role in fixing the size of the Mott or charge transfer gap is an interesting open question.

\subsection{Relationship between one-band and three-band model}

We now consider the long-debated question to what extent the one-band
model is able to represent the electronic structure of the three-band model
and whether the one-band model is sufficient to describe the low-energy
physics. 
The Wannier orbitals in the one-band model are extended objects with both $p$- and $d$-character. 
Hence, the lower Hubbard band in this model should not be considered as simply a $d^{8}$ state, but rather as
a representative of the Zhang-Rice and lower Hubbard bands found in the three-band calculation. 
Conversely, the upper Hubbard band in the one-band case is not simply a $d^{10}$ state, but an excitation which 
has no simple correspondence in the three-band calculation. 
If we consider the Zhang-Rice singlet band a substructure of the lower Hubbard band, we should compare the separation between the
Hubbard bands in the one-band calculation to the separation between upper Hubbard band and the Zhang-Rice band in the
three-orbital model, rather than to the 7 eV gap between the low-energy hump in the lower Hubbard band and the upper Hubbard band. 
In this case,
the agreement between the spectra seems acceptable, given the difference in localization between the Wannier orbitals. 

Of course, the one-band calculation cannot reproduce the strong asymmetry of the three-orbital model $d$-spectral function, which 
originates from the presence of the oxygen bands. Also, the gap size is too small, since the calculation does not take into account
the effect of $U_{pd}$, which is essential in fixing the $p$-$d$ level splitting
in the three-band calculation.

\section{Summary}
\label{summary}

We have constructed low-energy one-band and three-band
models for La$_{2}$CuO$_{4}$ from first-principles. The one-particle band structure was based
on the LDA and the frequency-dependent effective interaction (dynamic $U$) was 
calculated using the cRPA method. In both models LDA+DMFT calculations
using a static $U$ taken as the zero frequency limit of the dynamic $U$ do not
yield the expected insulating gap. It is necessary to take into account the
frequency-dependent $U$ in order to open up a gap in the spectrum. This
clearly shows the crucial role of dynamical sceening in a correct description of 
the insulating state of La$_{2}$CuO$_{4}$ and in
obtaining a consistent picture of the low-energy electronic structure. 
In agreement with Ref.~\onlinecite{Hansmann2013} we have
also found that it is important to take into account the change in the inter-atomic 
Hartree potential, which is neglected in most DMFT
calculations, to get the correct position of the oxygen $p$ band relative to
the $d$ band.

We found that the $d$ states which should be identified with the lower
Hubbard band cover the same energy range as the $p$ states, and that the Zhang-Rice band should be considered
a substructure of the lower Hubbard band.  In addition two pronounced collective excitations embodied in the fully
screened interaction $W$ were observed  at $\omega=3$ and $9$ eV. The peak at
$3$ eV can be traced back to a collective plasmon-like excitation arising from
particle-hole excitations within the antibonding $d$ band whereas the $9$ eV
peak corresponds to a collective excitation originating from transitions
between the occupied oxygen $p$ bands and the antibonding $d$ band. The peak
at $9$ eV is responsible for the very strong screening effect in La$_2$CuO$4$. It also 
gives rise to satellites in the spectral function at
$-13$ and $+10$ eV. The peak at $-13$ eV that may look like the lower
Hubbard band is in fact a subplasmon satellite associated with the $p$-to-$d$
transitions. The true lower Hubbard band is partially masked by the oxygen $p$ bands
at a lower binding energy.

Comparison between the spectral functions of the one- and three-band models
reveals that the one-band model is not sufficient to describe the electronic
structure within the energy range of the gap. The size of the gap of the
one-band model is significantly smaller than that of the three-band model, where the
latter value of $1.9$ eV is in very good agreement with the experimentally
measured data of $2.0$ eV. It is also quite evident that the one-band model
cannot properly describe the true character of the top of the valence band, which is of
the type $d^{8}$ ligand hole, rather than a simple lower Hubbard band
splitting off the antibonding $d$ band.

While our calculation is based on the {\it ab-initio} bandstructure, uses {\it ab-initio} interaction parameters, 
and takes into account the $p$-$d$ interaction (at the Hartree level), 
one missing ingredient is the momentum-dependence of the self-energy. 
It has been found in recent studies that the strong band renormalization from the dynamical $U$ is
at least partially compensated by a band-widening due to the $k$-depdendence of 
the self-energy.\cite{Miyake2013,Sakuma2013,Roekeghem2014} 
Quantifying these effects for La$_2$CuO$_4$ requires more advanced schemes,
such as cluster extensions of DMFT (which cannot be easily combined with the most efficient techniques for treating 
frequency dependent $U$), or GW+DMFT (which may not
properly capture the $k$-dependence in strongly correlated compounds\cite{Ayral2013}). 
Exploring these issues will be an interesting topic for future studies. 

\acknowledgments

We thank P. Hansmann for helpful discussions. The DMFT simulations have been run on the Brutus cluster at ETH Zurich, using a code based on ALPS.\cite{ALPS}
PW has been supported by SNSF grant No.~200021-140648 and NCCR Marvel. Some of this research has been carried out during a stay at ESI (Vienna). 
This work was also supported by the Swedish Research Council and part of the
computations were performed on resources provided by the Swedish National
Infrastructure for Computing (SNIC) at LUNARC.


\begin{thebibliography}{99}
\bibitem{DC}{Use of the LDA density in the double counting term for the $d$ band would push the $d$ bands down into the $p$ bands and result in a metallic solution. This is because $n_d^\text{LDA}\approx 1.5$ is considerably higher than the $n_d\approx 1.1$ of the insulating solution.}
\end{thebibliography}
\end{document}